\documentclass[aps,pre,reprint,showpacs,superscriptaddress]{revtex4-1}
\usepackage{graphicx}
\usepackage{dcolumn}
\usepackage{bm}
\usepackage{color}
\usepackage{lipsum}
\usepackage{mathrsfs}
\usepackage{lineno}
\usepackage[colorlinks,urlcolor=blue,linkcolor=blue,citecolor=blue]{hyperref}
\usepackage{siunitx}
\usepackage{multirow}
\begin{document}


\title{Pinch effect of self-generated magnetic fields in the quantum degenerate plasmas on the heating process of the double-cone ignition scheme}
\author{Y.-H. Li}
\affiliation{Institute of Physics, Chinese Academy of Sciences, Beijing 100190, People’s Republic of China}
\affiliation{University of Chinese Academy of Sciences, Beijing 100049, People’s Republic of China}
\affiliation{Collaborative Innovation Center of IFSA (CICIFSA), Shanghai Jiao Tong University, Shanghai 200240, People’s Republic of China}
\author{D. Wu}
\email{dwu.phys@sjtu.edu.cn}
\affiliation{Key Laboratory for Laser Plasmas and Department of Physics and Astronomy, Shanghai Jiao Tong University, Shanghai 200240, People’s Republic of China}
\affiliation{Collaborative Innovation Center of IFSA (CICIFSA), Shanghai Jiao Tong University, Shanghai 200240, People’s Republic of China}
\author{J. Zhang}
\email{jzhang1@sjtu.edu.cn}
\affiliation{Institute of Physics, Chinese Academy of Sciences, Beijing 100190, People’s Republic of China}
\affiliation{University of Chinese Academy of Sciences, Beijing 100049, People’s Republic of China}
\affiliation{Key Laboratory for Laser Plasmas and Department of Physics and Astronomy, Shanghai Jiao Tong University, Shanghai 200240, People’s Republic of China}
\affiliation{Collaborative Innovation Center of IFSA (CICIFSA), Shanghai Jiao Tong University, Shanghai 200240, People’s Republic of China}

\date{\today}
\begin{abstract}
In the double-cone ignition scheme, compressed fuels in two head-on cones are ejected to collide, forming a colliding plasma with an isochoric distribution for rapid heating by high flux fast electrons from picosecond petawatt laser beams in the perpendicular direction from the cone axis. In this work, we investigate the effects of quantum degeneracy on the transport of fast electrons in the colliding plasma, which rapidly evolves from the quantum degenerate in the outer region of the plasma to the classical state in the concentric core region heated by the colliding fronts of the plasma jets. With large scale particle-in-cell simulations, it is found that the self-generated magnetic field generated by the transport of fast electrons in the quantum degenerate state at the outer region is much stronger than in the corresponding classical state with the same fuel density in the core region. Theoretical analysis of the growth of the self-generated magnetic field is developed to explain the simulation results. Such strong self-generated magnetic fields in the quantum degenerate states can pinch the axially injected fast electrons to deposit their energy in the concentric core region, improving the heating eﬀiciency for fast ignition.
\end{abstract}


\maketitle

\section{Introduction}
In the double-cone ignition (DCI) scheme \cite{dci}, the Deuterium-Tritium fuel shells are separately placed in two head-on cones. Under the confinement and guidance of cones, the two shells are compressed and accelerated by carefully tailored nanosecond laser pulses to eject from the cone-tips and finally collide head-to-head, forming a concentrically pre-heated plasma with extremely high density. Then fast electron beams produced by picosecond petawatt laser pulses are injected along the direction perpendicular to the compression cone axis into the preheated fuel, deposit their energy rapidly, creating a hot spot in a very localized region of the fuel. Compared with the central ignition scheme \cite{nuckolls1972laser}, the DCI scheme significantly reduces the energy requirement for both the compression and heating laser pulses, as well as relaxes the tolerance to compression asymmetry and hydrodynamic instabilities.

In the fast heating process of the DCI scheme, the transport and deposition of fast electrons in the pre-heated fuel plasma make an essential difference to the quality of hot spot. Different from the fast ignition (FI) scheme \cite{tabak1994ignition}, fast electrons in the DCI scheme are injected into a pre-heated isochoric plasma, formed by the colliding of two degenerated plasma jets from the tips of the cones. In the colliding processes, kinetic energy of the plasma jets are converted into thermal energies. Therefore, rapid phase transition between quantum degenerate and classical states takes place in the colliding front. By carefully modulating the collision velocity and other factors, e.g., the injection timing of fast electrons, the state of degeneracy of the fuel is therefore tunable, which provides great opportunities in order to optimize the transport and deposition of fast electrons.

For conventional FI scheme, the fast heating process remains a bothering issue. On the one hand, the heating electron beam generated by petawatt laser pulse has a wide energy spectrum, for which reason the high-energy part of electrons penetrates through the ideal hot spot region; On the other hand, the angular diffusion of electrons also disperses the deposition of energy, hence hindering the efficiency of heating. A large amount of important work has been conducted on the fast heating process in previous FI researches. Kodama et al carried out early experiments at Gekko  $\si{\uppercase\expandafter{\romannumeral13}}$ laser facility \cite{kodama2001fast,kodama2002fast}, observing a more than 20$\%$ heating efficiency. However, subsequent studies \cite{azechi2013present,arikawa2017improvement,kitagawa2022direct} have not repeated such an optimistic result. At the same time, the self-focusing of electron beam triggered by macroscopic magnetic field was attached great importance \cite{davies2003electric,robinson2012focusing,ning2021self}. To step further on the heating efficiency, the magnetically assisted fast ignition scheme had been developed by Wang et al \cite{wang2015magnetically} to constrain the electron beam by implicating an external magnetic field along the beam. In terms of numerical simulation, hybrid PIC model \cite{davies2002wrong,gremillet2002filamented} was developed to precisely describe electron transport in a full-scale configuration. In particular, it treats the injecting relativistic electrons by PIC while describes the background by the simple Ohm’s law. Additionally, the Monte Carlo method was introduced into PIC simulations, describing collision between particles in extreme dense plasmas \cite{sentoku2008numerical,mishra2013collisional}. 

Though the study of electron transport is well established, the quantum degeneracy effect has not yet taken seriously, which would play important roles in DCI scheme. Recently researchers of National Ignition Facility \cite{hayes2020plasma} has also confirmed the electron degeneracy in the compressed cold D-T fuel in their central ignition experiments, but no conclusion has been drawn so far whether degeneracy exerts a positive influence to the formation of hot spot. 

For this problem, we have carried out detailed simulation research and made theoretical explanations to our result. It reveals that a strong magnetic field is generated when fast electrons transport in degenerate plasmas, which pinches injected electrons to deposit energy in localized region, which might be beneficial for improving the efficiency of fast electron heating for ignition. 
This finding is useful for advanced fast ignition scheme to improve the heating efficiency by modulating the state of degeneracy of D-T fuel. 

This paper is organized as follows. Section $\si{\uppercase\expandafter{\romannumeral 2}}$ shows the simulation parameters and result of monoenergetic electron transport in degenerate and non-degenerate D-T plasmas with LAPINS code. Section $\si{\uppercase\expandafter{\romannumeral 3}}$ gives theoretical analysis to the simulation results. In Section $\si{\uppercase\expandafter{\romannumeral 4}}$, our conclusions are finally presented.

\section{Simulation results}
To distinguish the prime difference between degenerate and non-degenerate plasma for electron depositions, we use LAPINS \cite{wu2017monte,wu2019high,wu2020particle,wu2019particle} code to conduct 2D3V simulations of monoenergetic electrons in uniform plasmas. LAPINS is compatible with calculation of degenerate and dense plasmas over large scale by improving interactions of collision and electromagnetic field. Compared with conventional PIC codes, LAPINS modifies the kinetic equation to Boltzmann-Uhling-Uhlenbeck equations, where the Pauli exclusion principle is considered in collision term. With this approach \cite{wu2020particle}, Fermi-Dirac statics are implemented to degenerate particles. Under the circumstance that density of fast electrons is several orders smaller than that of background electrons, electrons in background plasma are treated as fluid \cite{wu2019particle}, solved by alternative simultaneous equations of Ampere’s law, Faraday’s law and Ohm’s law.

The simulated configuration is shown as Table\ \ref{table1}. In the simulation, the particle number density is normalized by $n_c=1.16\times 10^{21} \ \si{cm^{-3}}$. The $60\ \si{\mu m} \times 40\ \si{\mu m} $ simulation region on $yOz$ plain is filled with uniform D-T plasma, whose number density is set to be $n_e=30000n_c$ (approximately $139 \ \si{g}/\si{cc}$), corresponding to a Fermi temperature of $379.33\ \si{eV}$. The temperature of plasma is respectively $50\ \si{eV}$ and $500\ \si{eV}$ on the either side of Fermi temperature, representing the degenerate and non-degenerate state. A $5\ \si{MeV}$ monoenergetic electron beam is injected along $z$ axis, represented as an electron source at $z=2.5\ \si{\mu m}$ emitting fast electrons towards the positive $z-axis$ direction in the simulations. The beam number density follows a normal distribution $n_b(y)=n_b(0)\exp\left(-y^2/2\sigma^2\right)$ along y axis, with a peak value of $n_b(0)=0.05n_c $ in the center and a waist of $ \sigma=10\ \si{\mu m}$.

\begin{figure*}
\includegraphics[width=17cm]{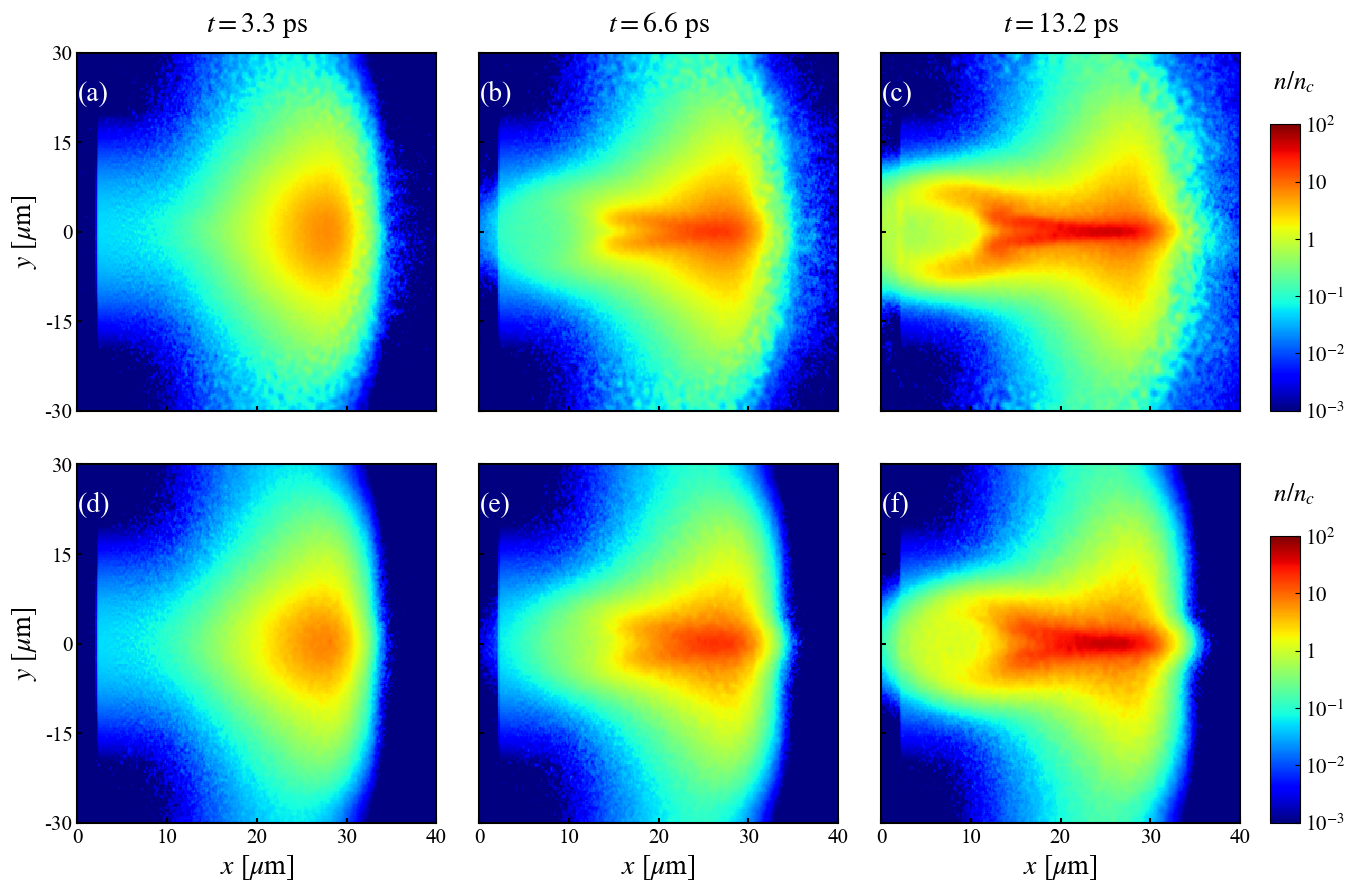}
\caption{\label{fig1}(color online). Snapshots of normalized number density of injected $5\ \si{MeV}$ fast electrons $n_b/n_c$ in initially degenerate (a-c) and non-degenerate (d-f) D-T plasma.}
\end{figure*}

\begin{table}[b]
\caption{\label{table1}%
Simulation configurations of electron transport, sequentially listing the number density of background electrons and fast electrons (in unit of $n_c=1.16\times 10^{21} \ \si{cm^{-3}}$), the kinetic energy of fast electrons, and the initial temperature of plasmas.
}
\begin{ruledtabular}
\begin{tabular}{rcccc}
\textrm{Scheme}&
$n_e$ ($n_c$)&
$n_b$ ($n_c$)&
$E_k$ ($\si{MeV})$&
$T$ ($\si{eV}$)\\
\colrule
Degenerate & \multirow{2}*{30000} & \multirow{2}*{0.05} & \multirow{2}*{5} & 50\\
Non-degenerate & ~ & ~ & ~ & 500\\
\end{tabular}
\end{ruledtabular}
\end{table}
Fig.\ \ref{fig1} shows the number density of injected electrons. The electrons mostly deposit at $z=35\ \si{\mu m}$.  The three frames of pseudo-color images are respectively caught at $3.3\ \si{ps}$, $6.6\ \si{ps}$, and $13.2\ \si{ps}$. It comes out that electrons gathered as a spherical bulk in the early stage of both simulations, while the core of deposit region (above    $10n_c$) is pinched to a long strip over time. Additionally, two low density tails (mostly less than  $n_c$) grow up behind the deposit region. It is noted that such processes take place faster and more drastic in the initially degenerate plasma.

\begin{figure*}
\includegraphics[width=17cm]{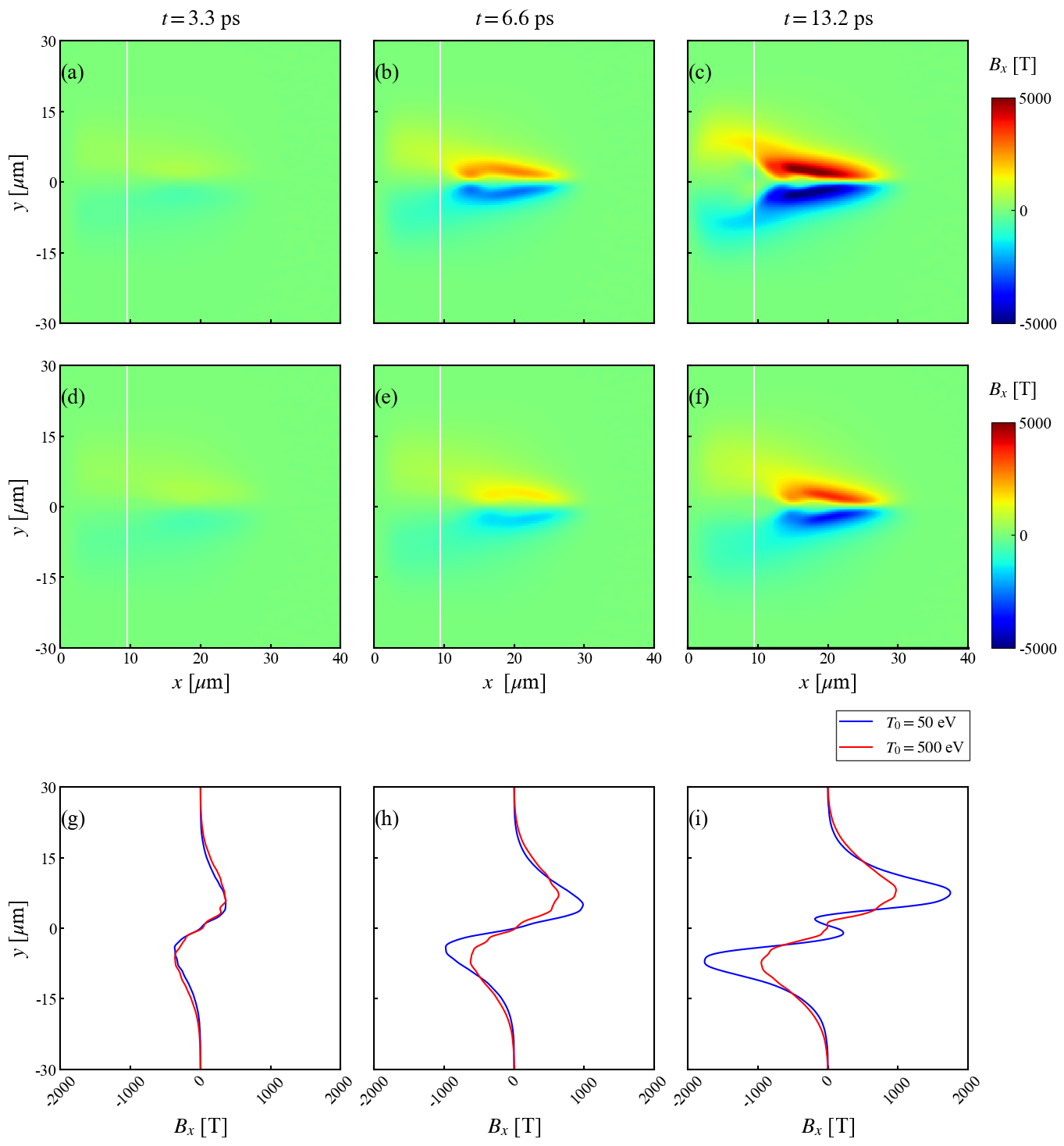}
\caption{\label{fig2}(color online). Snapshots of magnetic field $B_x$ ($\si{T}$) in initially degenerate (a-c) and non-degenerate (d-f) D-T plasma, with 1-D section curves (g-i) at $z = 9.5\ \si{\mu m}$ along $y$ axis in the bottom row.}
\end{figure*}

These phenomena are considered to be aroused by the self-generated electromagnetic field in plasma. As shown in Fig.\ \ref{fig2}(a-f), a contracting magnetic field of thousands of Teslas is yielded along x axis. This field is triggered by the return current of plasma following Faraday’s law $\partial\boldsymbol{B}/\partial t=-\boldsymbol{\nabla} \times(\eta \boldsymbol{J_b})$, where $\eta$ is the resistivity of plasma, and $\boldsymbol{J_b}$ is the density of return current caused by background drifting electrons. The field is strong enough that the magnetic pressure overwhelms the thermal pressure and pinches the electron bunch. A stronger return current is excited and in turn multiplies the growth of the magnetic field. Hence, the magnetic field is several times stronger over $z=20\ \si{\mu m}$ than that at the edge of the plasma, and an obvious focusing trend is spotted. Under the same color bar, it is intuitive to find that the magnetic field in the initially degenerate plasma is much stronger, therefore the pinch effect on electrons is more significant. Fig.\ \ref{fig2}(g-i) depict the profile of magnetic field along y axis at $z=9.5\ \si{\mu m}$. There is a peak couple recognized as the cross section of a toroidal magnetic field before $6.6 \ \si{ps}$, while the field begins to flip from the center afterwards. An evident opposite pair of peaks is observed in Fig.\ \ref{fig2}(i), forming a concentric configuration with the original one. From a three-dimensional perspective, the electrons between two repelling fields are squeezed as a barrel, corresponding to the low density tails shown in Fig.\ \ref{fig1}. 

\begin{figure}
\includegraphics[width=8.5cm]{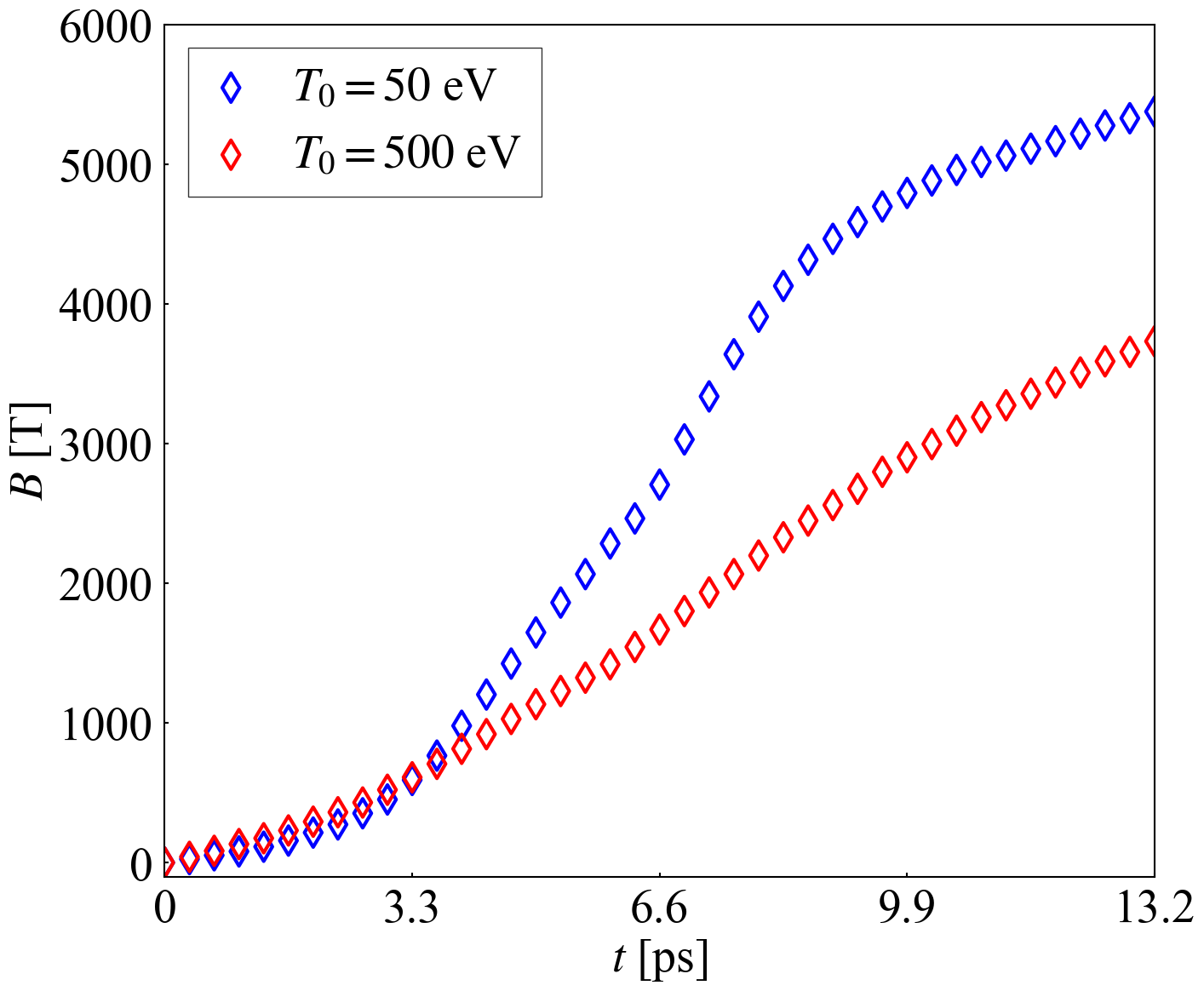}
\caption{\label{fig3}(color online). Maximum value of magnetic field $B_x$ ($\si{T}$) in D-T plasmas over simulation time.}
\end{figure}

Fig.\ \ref{fig3} quantitatively shows the maximum value of the magnetic field intensity in the plasma over time. The blue and red lines representing the field in the initially degenerate and non-degenerate plasma respectively. At the beginning, the field of degenerate plasma is weaker, due to the low resistivity in the degenerate state. As plasma is heated, the blue line grows exponentially, and its value rapidly becomes nearly twice as much as that of the red line. After $10\ \si{ps}$, the magnetic field in two simulations tend to saturate successively, indicating that the electron thermal pressure turns comparable to the magnetic pressure and pinch effect is no longer dominant in the growth of the magnetic field.

\section{Theoretical analysis of self-generated magnetic field}
In this section a fundamental theoretical analysis of self-generated magnetic field in the electron transport is displayed. This study is mainly devoted to the influence of quantum effect, excluding the subsequent magnetic focusing process, since that stronger magnetic field results in greater pinch effect is self-evident. Two assumptions are made in the following calculation: (1) The current density $\boldsymbol{J_b}$ is constant over time; (2) The heat conduction is out of consideration, which means the temperature change caused by energy deposition is localized.
Consider that a monoenergetic electron beam with the current density of $\boldsymbol{J_b}=n_b e\boldsymbol{v_b}$ is injected into the D-T plasma along $z$ axis, where $\boldsymbol{v_b}=v_b \boldsymbol{\hat{e}}_z$ and $\boldsymbol{J_b}=J_b \boldsymbol{\hat{e}}_z$. To maintain macroscopic electrical neutrality, electrons in the background plasma are aroused to drift backwards as $ \boldsymbol{J_e}=-\boldsymbol{J_b}$. An electric field equals to resistivity times current density, $\boldsymbol{E}=\eta \boldsymbol{J_e}=-\eta \boldsymbol{J_b}$, is generated to decelerate fast electrons. If $\boldsymbol{E}$ is not spatially uniform, according to Faraday’s law, the growth rate of magnetic field component along $x$ axis is written as

\begin{equation}
    \frac{\partial B_x}{\partial t}=\frac{\partial}{\partial y}\left(\eta J_b\right)
    =\frac{\partial J_b}{\partial y} \left(\eta +J_b \frac{\mathrm{d} \eta}{\mathrm{d} J_b} \right),
    \label{1}
\end{equation}

\noindent where the derivative of $\eta$ with respect to $J_b$ is only mathematically, to temporarily treat $J_b$ as an independent variable in the explicit expression of $\eta$, while the current density in actual physical process is constant.

According to Eq.\ (\ref{1}), the growth of the magnetic field is closely dependent on $\eta$. When the number density of background electrons $n_e$ is much smaller than that of injected electrons $n_b$, $n_e$ is considered to be constant, and $\eta$ is only related to the temperature $T$. Using the chain rule, $\mathrm{d}\eta/\mathrm{d}J_b$ equals to $(\mathrm{d}\eta /\mathrm{d}T)(\partial T/ \partial J_b )$. In Davies’ pioneering work \cite{davies2003electric}, $\eta$  is set to be proportional to $T^\alpha$ with an arbitrary factor $\alpha$, and $ \mathrm{d}{\eta}/\mathrm{d}T=\alpha \eta/T$; To express $\partial T/ \partial J_b$, the relationship between T and $J_b$ is derived by solving

\begin{equation}
    c_v\frac{\partial T}{\partial t}=P_{heat},
    \label{2}
\end{equation}

\noindent where $c_v$ is the heat capacity and $P_{heat}$ is the heating power of electron beam. Davies’ calculation (see Ref. \cite{davies2003electric} for details) has already reflected the exponential growth of the magnetic field in strong degenerate plasma with $\eta\propto T$, as well as the polynomial time-related growth rate for classic case with $\eta \propto T^{-3/2}$. However, its $P_{heat}$ leaves collisional heating component out, which turns out to dominate the heating process in dense plasma of $\sim100\ \si{g}/\si{cc}$. The rest of this section displays a more precise analysis for DCI regime and a numerical result of magnetic field.

Firstly, consider the right side of Eq.\ (\ref{2}). The heating energy all comes from injected electrons. In a short interval $\Delta t$, the injected electrons move forward $v_b \Delta t$. If each single electron losses $\Delta \varepsilon$ of energy from $z_0$ to $z_0+v_b \Delta t$ on average, then

\begin{equation}
    P_{heat} = n_b v_b \Delta \varepsilon = -\frac{J_b}{e} \frac{\mathrm{d}\varepsilon}{\mathrm{d}z}.
    \label{3}
\end{equation}

$\mathrm{d}\varepsilon/\mathrm{d} z$ is called the stopping power \cite{solodov2008stopping}, consisting of the ohmic component and the collisional component. The ohmic component $\mathrm{d}\varepsilon_o/\mathrm{d} z= e \eta J_b$ represents the deceleration effect of the electric field of return current $\boldsymbol{E}=-\eta \boldsymbol{J_b}$. Meanwhile, the injected fast electrons collide with background electrons, also transforming directional kinetic energy into disordered thermal energy, which corresponds to the collisional component $\mathrm{d} \varepsilon_c/\mathrm{d} z= \kappa_0 n_i [Z_i \ln {\Lambda_f}+ (Z-Z_i) \ln{\Lambda_b}]$. $Z_i$ and $(Z-Z_i)$ are respectively the number of free and bound electrons, $\ln{\Lambda_f}$ and $\ln{\Lambda_b}$ are respectively the Coulomb logarithms for collision with free and bounded electrons, and $\kappa_0 =4 \pi e^4 /m_e v_b^2$ is the stopping factor. For fully ionized D-T plasma $Z_i = 1$ and $(Z-Z_i )=0$. The Coulomb logarithm $\ln{\Lambda_f}$ is taken as $\ln{(1+\lambda_D/b)}$,where $\lambda_D = \sqrt{(T/4\pi n_e)(1+\beta^2/v_{th}^2)}$ is the Debye length and $b= \min\{e^2/{m_e \beta^2},\ \hbar /{m_e\beta}\}$ is the minimum distance between the two colliding particles, where $\beta$ is the reduced colliding velocity and $v_{th}=\sqrt{(3k_B T/2m_e )}$ is the thermal velocity of background electrons under temperature T. In DCI scheme, fast electrons have an energy of $\varepsilon \sim\ \si{MeV}$, therefore a relativistic relationship $v_b=c/\Gamma(\varepsilon)$ is added to stopping power as

\begin{eqnarray}
    \frac{\mathrm{d}\varepsilon}{\mathrm{d}z}&=&\frac{\mathrm{d}\varepsilon_c}{\mathrm{d}z}+\frac{\mathrm{d}\varepsilon_o}{\mathrm{d}z} \nonumber \\ 
    &=&-\frac{4\pi e^4 n_e}{m_e c^2}\Gamma(\varepsilon)\ln{\Lambda_f}-\eta e^2 n_b c \frac{1}{\sqrt{\Gamma(\varepsilon)}}  ,
    \label{4}
\end{eqnarray}

\noindent where $\gamma=1+\varepsilon/m_e c^2$ is the relativistic factor and $\Gamma(\varepsilon)=\gamma^2/(\gamma^2-1)$.

\begin{figure}
\includegraphics[width=8.5cm]{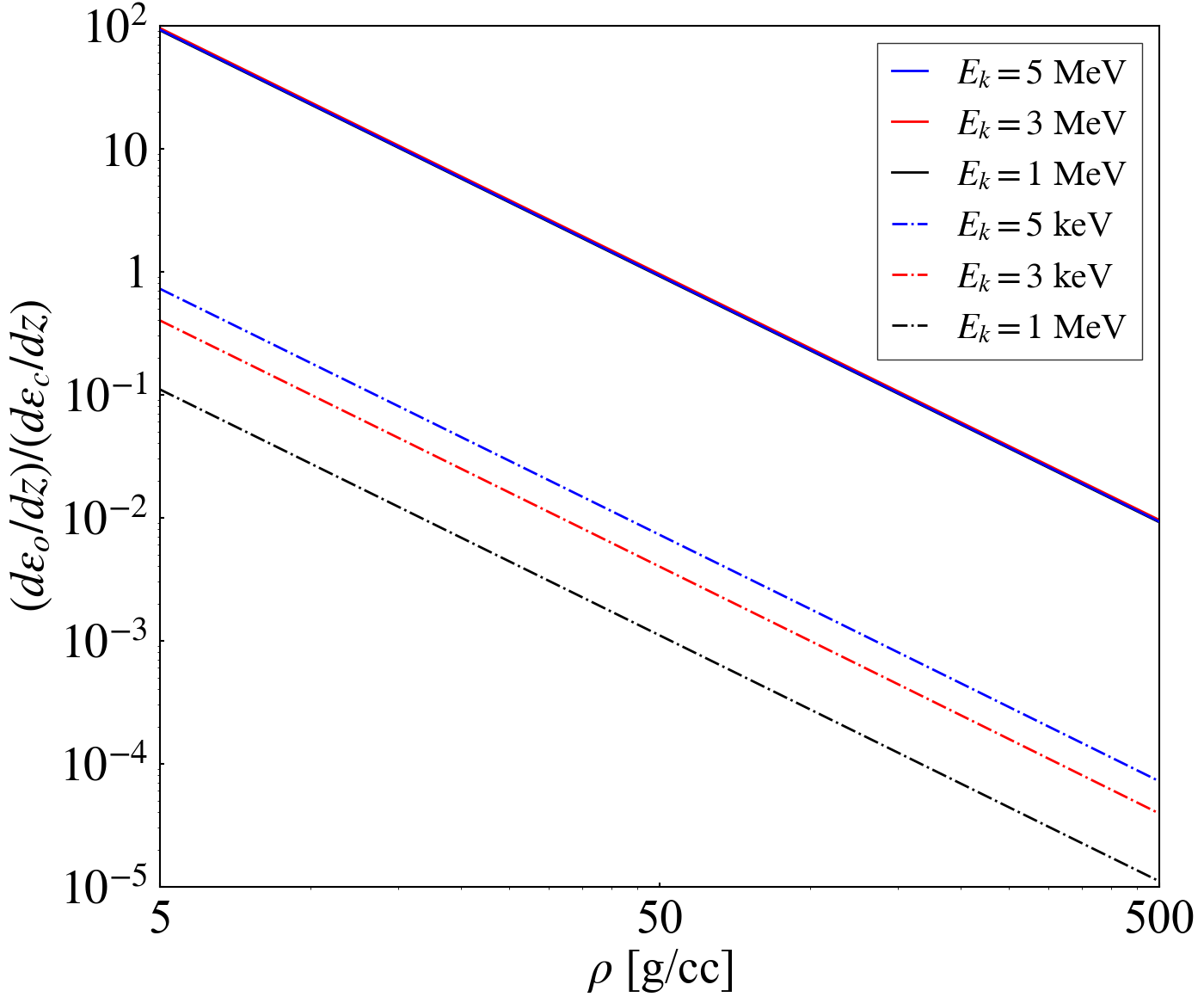}
\caption{\label{fig4}(color online). Ratio of the ohmic component $(\mathrm{d}\varepsilon_o/\mathrm{d}z)$ to the collision component $(\mathrm{d}\varepsilon_c/\mathrm{d}z)$ as a function of background plasma's density $\rho$ ($\si{g}/\si{cc}$), under the corresponding Fermi temperature of each certain $\rho$.} 
\end{figure}

For our interest, the ratio of ohmic component to collisional component of $\sim \ \si{MeV}$ and $\sim \ \si{keV}$ fast electron is calculated. The density of plasma $\rho$ ranges from $5\ \si{g}/\si{cc}$ to $500\ \si{g}/\si{cc}$, with the temperature set to be the corresponding Fermi temperature of each density. As depicted in Fig.\ \ref{fig4}, the collisional component can be neglected only in low density regime, for example, when electrons transport in Al with $\rho=2.7\ \si{g}/\si{cc}$. In the regime of plasma for DCI scheme to ignite with $\rho \sim 100\ \si{g}/\si{cc}$, the ratio $(\mathrm{d}\varepsilon_o/\mathrm{d}z)/(\mathrm{d}\varepsilon_c/\mathrm{d}z)$ falls below $0.1$, meaning that the collisional heating far exceeds the ohmic heating. 

The heat capacity $c_v$ is consists of two parts, the ion heat capacity and the electron heat capacity. Both heat capacity equal to classic value ${3k_B n_e}/2$ when $T$ is high enough. In the strongly degenerate state, it can be proved that the electron heat capacity is proportional to $T$, based on the Taylor expansion of FD function at zero temperature. Taking Fermi temperature $T_F$ as a turning point, $c_v$ approximates to a piecewise function
 
\begin{equation}
    c_v=\min\{3k_B n_e,\ \frac{3k_B n_e}{2}(1+T/T_F)\},
    \label{5}
\end{equation} 

As the crucial factor to determine the growth rate of the magnetic field, $\eta$ is also refined. It is calculated from Drude model $\eta=m_e \nu_e/e^2 n_e$, where $\nu_e$ is the total electron collision frequency. In hot regime $T\gg T_F,$ the classic Spitzer formulation $\nu_e={3e^4 n_e \ln\Lambda_f}/4\sqrt{2} m_e^{1/2} (k_B T)^{3/2}$ is applicable, and $\eta$ is proportional to $T^{-3/2}$. However, in the degenerate regime $T\ll T_F$ where the Pauli exclusion principle inhibits the electron collisions, $\nu_e$ is proportional to $T$ and drops to zero at zero temperature. Considering both cases,  $\eta$ is written as: 

\begin{equation}
    \eta=\frac{3m_e^{1/2}e^2\ln{\Lambda_f}}
    {4\sqrt{2} k_B^{3/2}(T+T_F)^{3/2}} \min\{1,\ T/T_F\}.
    \label{6}
\end{equation}  
 
\noindent $(T+T_F)$ in the denominator compromises on the weak degenerate state. When $T\gg T_F$ Eq.\ (\ref{6}) is reduced to classic formulation; when $T \ll T_F$, the denominator turns to be $(k_B T_F )^{3/2}$ neglecting the contribution from $T$. 
 
Now all the pre-works have been done for numerically solving the magnetic field.  substitute Eq.\ (\ref{3}), (\ref{4}), (\ref{5}) and (\ref{6}) into Eq.\ (\ref{2}), and figure out $T$ dependent on $t$ and $J_b$. Rouge-Kutta method is recommended for the deduction here. Then, $\partial T/\partial J_b$ is obtained. Substitute this time-varying $T$ back into Eq.\ (\ref{6}) to calculate $\eta$ at specific moment, as well as its derivative to temperature $\mathrm{d} \eta/\mathrm{d} T$. Taking $\eta$, $\mathrm{d} \eta/\mathrm{d} T $ and $\partial T/\partial J_b$ into Eq.\ (\ref{1}) and integrating both side over time, the magnetic field $B_x$ is finally solved out.

Specific parameters used for calculation are as follows. The density of plasma is set to be $140\ \si{g}/\si{cc}$ consistent with the simulation setting, corresponding to $n_e=3.35\times10^{25} \ \si{{cm}^{-3}}$. Temperature for two regimes is respectively $50 \ \si{eV}$ and $500 \ \si{eV}$. The number density of $5 \ \si{MeV}$ electron beam is $n_b=5\times10^{19} \ \si{{cm}^{-3}}$. $\partial J_b/\partial y$ is taken at $y=\sigma$ for Gauss distribution $J_b (y)=J_b (0)\exp(-y^2/2\sigma^2)$ with $\sigma=10\ \si{\mu m}$, approximating to $\partial J_b/\partial y=0.6J_b/\sigma$.

\begin{figure}
\includegraphics[width=8.5cm]{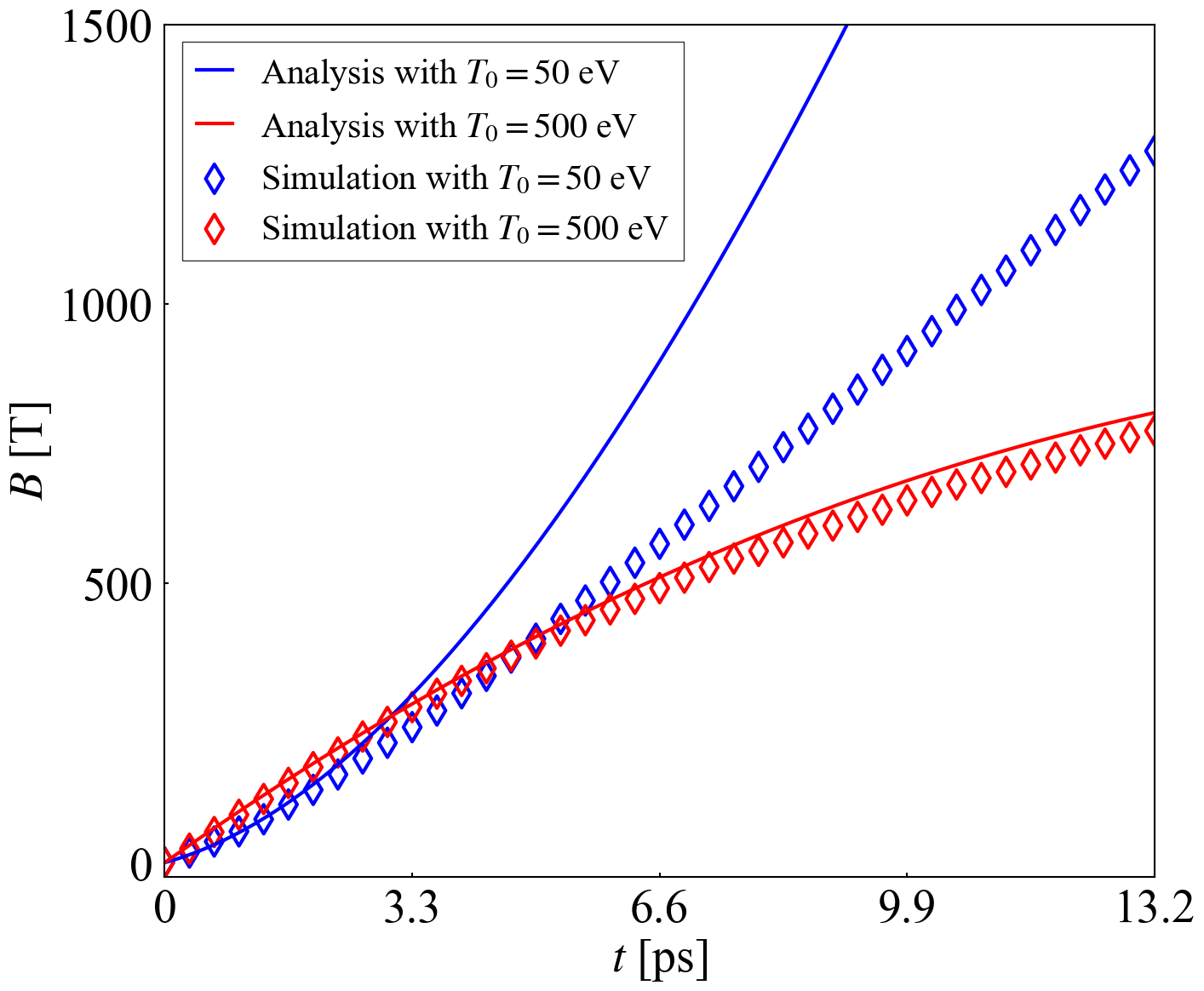}
\caption{\label{fig5}(color online). Theoretically calculated growth of magnetic field $B_x$ ($\si{T}$) (lines), compared with simulation results picked at $y=\pm10\ \si{\mu m}$, $z=6.25\ \si{\mu m}$ (diamonds).}
\end{figure}

Fig.\ \ref{fig5} shows the numerical calculation of magnetic field in lines compared with the field read from the PIC simulations in diamond scatters. The data of the simulations is picked as the peak of $B_x$ at $z=6.25\ \si{\mu m}$ , $\left| y \right| = \sigma=10\ \si{\mu m}$, where the current density can be regarded as the initial value and magnetic focusing effect is not significant yet. For the case that fast electrons transport in initially $500\ \si{eV}$ non-degenerate DT plasma, the theoretically calculated $B_x$ is in great agreement with the PIC simulation results. As for the degenerate case, the blue line and scatters match with each other in the beginning $3.3\ \si{ps}$, while deviations occur with theoretically calculated $(B_x )_{cal}$ grows much faster.

In further checking, the estimation of heating power is considered to be the main source of error. When electron beam is injected into the plasma, an extra charge-separation field is generated due to the polarization effect. This field is not concerned in the stopping power introduced above. Moreover, in PIC simulations, the number density of fast electrons $n_b$ in the plasma is found to be several times higher than the initial settings over time, though the current density $J_b$ remains steady. In other words, the beam velocity $v_b$ is actually decreasing, and according to Eq.\ (\ref{4}) the collisional component $\mathrm{d}\varepsilon_c/\mathrm{d}z$ is underestimated in the calculation. Consequently, the temperature $T_{sim}$ in simulation increases by several hundreds of electron volts, while $T_{cal}$ in calculation increases by merely around $100\ \si{eV}$. Just take the degenerate state for discussion. $T_{sim}$ has approached to the turning point $T_F$ at $t>10ps$, and the growth rate tends to flatten out; However, $T_{cal}$ is still in strong degenerate regime and $(B_x)_{cal}$ keeps rising exponentially. The Faraday’s law Eq.\ (\ref{1}) is unmistakable, so the theoretical calculation makes the correct prediction at the beginning, and the accumulation of error in temperature shows up later. As for non-degenerate case, it is possible that $\partial{B_x}/\partial t$ is not so sensitive to the error in high temperature regime, and deviations have not turned obvious in $13.2\ \si{ps}$ duration of the simulation.

\section{Conclusion and discussions}

In this paper we have investigated the effects of quantum degeneracy on the electron transport in the rapid heating process in the DCI scheme. Using PIC simulations, we find that a nearly doubled strong magnetic field is self-generated in an initially degenerate plasma, compared with the case in a non-degenerate plasma, leading to a strong pinch effect along the axis of the injected electrons via magnetic focusing process. We extended the Davies’ model, where collisional component is particularly taken into account, which plays a dominating role in the dense plasma. Such theoretical calculations lead to a  similar result of growth of the magnetic field with PIC simulations. 

This research has potential applications for fast ignition. It reveals that a strong magnetic field is generated when fast electrons are transported in a degenerate plasma, resulting in an efficient energy deposition of injected fast electrons in the pinch region. This would certainly be beneficial for improvement of the heating eﬀiciency of fast electrons in the DCI scheme.

There are still some issues for further improvement in this study. Firstly, the polarization effect is needed to take into consideration for a more precise description of the heating process; Secondly, the quantitive calculations of two-dimensional magnetic pinch effects are not included, and the maximum magnetic field in the plasma is still not fully investigated. Moreover, the flipping of magnetic field as shown in see Fig.\ \ref{fig2}(i) and its influence on electron deposition is observed in simulations. The role of magnetic flipping on electron transport and energy deposition still needs further investigations.

\begin{acknowledgments}
This work is supported by the Strategic Priority Research Program of Chinese Academy of Sciences (Grant Nos. XDA25010100 and XDA250050500), National Natural Science Foundation of China (Grants No. 12075204), and Shanghai Municipal Science and Technology Key Project (No. 22JC1401500). Dong Wu thanks the sponsorship from Yangyang Development Fund.

\end{acknowledgments}

\end{document}